\def\PR{{Phys.~Rev.~}}
\def\PRL{{ Phys.~Rev.~Lett.~}}

\def\sVEC{\text{\small{VEC}}}
\newcommand{\onlinecite}[1]{\hspace{-1 ex} \nocite{#1}\citenum{#1}}

\documentclass[aps,prb,showpacs,superscriptaddress,a4paper,twocolumn]{revtex4-1}

\usepackage{chngcntr}
\usepackage{enumerate}
\usepackage{amsmath}
\usepackage{color}
\usepackage{floatrow}
\usepackage[rightcaption]{sidecap}
\usepackage{amssymb}
\usepackage{graphicx}

\begin{document}

\title{Design of high-strength refractory complex solid-solution alloys}

\author{Prashant Singh}
\affiliation{Ames Laboratory, U.S. Department of Energy, Iowa State University, Ames, Iowa 50011 USA.}
\email{prashant@ameslab.gov} 

\author{Aayush Sharma}
\affiliation{Mechanical Engineering, Iowa State University, Ames, Iowa 50011 USA.}

\author{A. V. Smirnov}
\affiliation{Ames Laboratory, U.S. Department of Energy, Iowa State University, Ames, Iowa 50011 USA.}

\author{Mouhamad S. Diallo}
\affiliation{Mechanical Engineering, Iowa State University, Ames, Iowa 50011 USA.}

\author{Pratik K. Ray}
\affiliation{Ames Laboratory, U.S. Department of Energy, Iowa State University, Ames, Iowa 50011 USA.}
\affiliation{Materials Science \& Engineering, Iowa State University, Ames, Iowa 50011 USA.}

\author{Ganesh Balasubramanian}
\affiliation{Mechanical Engineering \& Mechanics, Lehigh University, Bethlehem, PA 18015 USA.}

\author{Duane D. Johnson}
\affiliation{Ames Laboratory, U.S. Department of Energy, Iowa State University, Ames, Iowa 50011 USA.}
\affiliation{Materials Science \& Engineering, Iowa State University, Ames, Iowa 50011 USA.}
\email{ddj@amelab.gov;ddj@iastate.edu}

\begin{abstract}
Nickel-based superalloys and near-equiatomic high-entropy alloys containing Molybdenum are known for higher temperature strength and corrosion resistance. Yet, complex solid-solution alloys offer a huge design space to tune for optimal properties at slightly reduced entropy.  For refractory Mo-W-Ta-Ti-Zr, we showcase KKR electronic-structure methods via the coherent-potential approximation to identify alloys over 5-dimensional design space with improved mechanical properties and necessary global (formation enthalpy) and local (short-range order) stability.   Deformation is modeled with classical molecular dynamic simulations, validated from our first-principles data. We predict complex solid-solution alloys of improved stability with greatly enhanced modulus of elasticity ($3\times$ at 300 K) over near-equiatomic cases, as validated experimentally, and with higher moduli above 500~K over commercial alloys ($2.3\times$ at 2000 K). We also show that optimal complex solid-solution alloys are not described well by classical potentials due to critical electronic effects.
\end{abstract}

\maketitle

\section*{Introduction}
\noindent    Nickel-based superalloys exhibit high-temperature strength, toughness, and oxidation resistance in harsh  environments.\cite{TMP2006} Improving existing single-crystal alloys is unlikely as melting is near 1350$^{\circ}$C, and, heat treatment lowers this to $\sim$1270$^{\circ}$C.  In high-speed turbines, melting reduces below 1250$^{\circ}$C  at the zone between the bond coat (e.g., NiAl) and the single-crystal blade.\cite{SJB2001} As such, the engine efficiency and thrust-to-weight ratio can be improved by a guided search for new materials. High-entropy alloys based on refractory elements may achieve higher temperature operation with superior creep strength.\cite{TZM} Typical refractory high-entropy alloys exhibit a yield strength of 500-700~MPa at 1200$^{\circ}$C, surpassing Ni-based superalloys.\cite{ONS2011} 
Indeed, at elevated temperatures Mo-based alloys show good thermal  (higher conductivities with lower strains\cite{Guo2015}) and mechanical  (machinability)\cite{P2009} properties, making them promising candidates.

\noindent    Almost all high-entropy alloys for which mechanical properties have been reported are based on Cr-Fe-Co-Ni with other elements added, e.g., Al,\cite{Yeh2004} Mn,\cite{Zhou2007} Mo,\cite{Kao2009} and Ti.\cite{Zhou12007} CoCrFeNi exhibits very high compression strength at 300 K, often exceeding 1500 MPa. Strains in as-cast condition do not often exceed 5$-$7\%, albeit a few exhibit 25$-$33\%.\cite{Zhou12007,Zang2010} Annealing does improve ductility of as-cast alloys.\cite{Zang2010} As with conventional alloys, a rapid decrease in strength (i.e., Young's modulus, E) occurs above $0.6$ of the melting temperature $\text{T}_{m}$, and the strength of alloys approaches 100 MPa at 1273 K.\cite{Yeh2004} 

\noindent  High-entropy alloys consist of $N$($\ge$5) elements in near equiatomic compositions ($c_{\alpha}$$\sim$$1/N$), giving maximal point (mixing) entropy ($S_{pt}$=$-\sum_{\alpha=1}^{N} c_{\alpha} \ln c_{\alpha} \xrightarrow{\text{max}} \ln N$),  that may better form solid solutions due to a compromise between the large $S_{pt}$ and a formation (mixing) energy $\Delta{E}_{form}$ that is not too positive (strongly clustering) nor too negative (strongly ordering).\cite{Yeh2004} 
As for binary solid solutions, Hume-Rothery's rules\cite{Hume-Rothery1969} for atomic size difference  ($\delta$), crystal structure,  valence electron concentration ($\sVEC$), and electronegativity difference ($\Delta\chi$) play a similar role in high-entropy alloy formation.
The production of several single or multi-phase alloys with face-centered cubic (FCC or {A1}), body-centered cubic (BCC or {A2}), hexagonal close-packed (HCP or {A3}), or cubic diamond ({A4}) structures exhibiting enhanced high-temperature strength, ductility, fracture and creep resistance to corrosion,\cite{Yeh2004,Zhou2007,Zhou12007,Kao2009,BG2014,ZL2016} and thermal stability\cite{VD2010} validates the concept of HEAs.\cite{Yeh2004} MoWVNbTa, e.g., with a density of 12.2 g/cm$^{3}$, has a reported usable strength up to 1873~K.\cite{ONS2011} 

\noindent    Nonetheless, from an alloy design perspective, complex solid-solution alloys (CSAs) offer a huge design space to tune properties, especially considering the strong effects alloying has on electronic properties (``band'' filling, hybridization, Fermi-surface nesting, ...), phase stability, and structure.  The CSAs comprised of whole composition (Gibbs) space, however, high-entropy alloys are a subset of it. Optimized CSAs offer a slightly reduced entropy with a single-phase region, or two-phase region for enhanced mechanical properties, existing in a desired operational temperature range.\cite{JohnsonCPA,JP1993,PS2015} 

\noindent  Here we narrow the design of high-strength, refractory (Mo-W)-Ta-(Ti-Zr) alloys via KKR electronic-structure methods within density-functional theory (DFT) using the coherent-potential approximation (CPA) to handle chemical disorder and thermodynamic averaging.\cite{JohnsonCPA,JP1993, PS2015,AS2016} The well-established KKR-CPA predicts structural properties [e.g., Young's (E) or bulk modulus (B)], and phase stability ($\Delta{E}_{form}$ vs. $\{c_{\alpha}\}$), as well as short-range order (SRO) via thermodynamic linear response,\cite{SJP1990,Pinski1991,SJP1994,AJPS1996,PS2015} a method, {in particular}, which revealed the origin for Hume-Rothery's size-effect rule.\cite{Pinski1991,Maddox1991} 
Notably, global stability ($\Delta{E}_{form}$) and local instability (SRO) should be jointly assessed:  While segregation is expected for $\Delta{E}_{form} > 0$, SRO can be segregating from local  compositional instabilities even if $\Delta{E}_{form} < 0$.
To  predict mechanical behavior (e.g., E)  versus temperature (T) rapidly, we performed extensive molecular dynamics (MD) simulations based on semi-empirical potentials, validated in part by first-principles results (and also highlighting limitations of such methods).
The tuned and proposed refractory quinary alloys and their properties are placed in context to Hume-Rothery-type design targets and compared to experiments.

\section*{RESULTS AND DISCUSSION}

\subsection{Hume-Rothery Design Targets:} High-entropy alloys contain elements with $c_{\alpha}$$\sim$35--12~at.\% ($N$=3-8). Trial-and-error has led to alloys with simple crystal structures, and a few with extraordinary properties,\cite{AdEngMat2004} e.g., formability using size disparate elements for confusion by design.\cite{Nat303} 
For CSAs design of phase stability and of electronic and mechanical behavior, targets for DFT-based KKR-CPA are limited by extending Hume-Rothery\cite{Hume-Rothery1969} criteria:
\begin{enumerate}
\small{
 \item {\bf Size:} Solute and host atomic radii (in elemental solid) must differ by $< 15\%$.\cite{Hume-Rothery1969,Pinski1991} For CSAs, with ${\bar{r}}$=$\sum_{i=1}^{N}  c_{i}{r_{i}}$, size limit in terms of standard deviation is sensible: \\ $0\le \delta \le 6\%$, 
with $\delta$ = $100\% $$\times$$ \left[\sum_{i=1}^{n}  c_{i}({r^2_{i}}-{\bar{r}^2})/{\bar{r}^2} \right]^{1/2}$.
\vspace{-2mm}
\item {\bf Lattice:} Similar crystal structures for solute and host. 
\vspace{-2mm}
\item  {\bf \sVEC:} Large solubility when solute and host have the same \sVEC.  
                     A metal dissolves one of higher (lower) valency to a greater (lesser) extent.
\vspace{-2mm}
\item  {\bf $\chi$'s:} If $\Delta\chi$ is too great, metals tend to form intermetallic compounds, not solid solutions.
\vspace{-2mm}
\item  {\bf $\Delta{E}_{form}$:} For $-11$$ \lesssim$$\Delta{E}_{form}$$\lesssim +5$~\text{mRy} CSAs stabilized in usable T's. 
}
\end{enumerate} 
{\par} A few  comments are warranted. 
In \#1, the 6-6.8\% achieves 15\%-rule for CSAs with $\sim$50\% confidence level, an inequality also found empirically.\cite{Tsai2010-MRL,Yang2012}
Extending \#3 via electronic density of states concepts, A2 forms for 4$<$\sVEC$<$6 as stability increases when bonding $d$-states fill, and is maximal when half-filled ($\sVEC$$\approx$6);  antibonding states fill with \sVEC$>$6 (above a pseudogap, see results)  and stability decreases. 
Indeed, A2 CSAs are observed when \sVEC~is 5$\pm$1.\cite{VEC1} For $6.8\le\sVEC\le8$ other phases compete, e.g., FeCr has $\sVEC$$=$7 (like Mn) and constituent's structure are both A2, yet the CSA is unstable to the $\sigma$-phase, as often appears.\cite{Tsai2010-MRL} 
Again, from band filling, A1 becomes more stable for \sVEC$>$8.\cite{Skriver1984}
CSAs are indeed observed to form within these rules.\cite{Zhang2014}
$\Delta{E}_{form}$ lower limit in \#5 is set by $-T_{a}S_{pt}$, where annealing temperature (needed for kinetics) is  T$_{a}$$\sim$0.55T$_{m}$ ($\sim$1000--1650~K for refractories); upper limit is set such that miscibility gap T$^{MG}_{c}$$<$T$_{a}$ (where 158~K$\sim$1~mRy). For $\delta>5\%$, CSAs with $\Delta{E}_{form}>5$~mRy form complex phases, but tend to form metastable metallic glasses for $\Delta{E}_{form}<-11$~mRy.\cite{Inoue1993}
Considering binaries and supercells, these limits for CSA formation are supported.\cite{2015PRX.HEA}  

\noindent   As we have shown, transition temperatures from $\alpha$ (CSA) to $\beta$ (ordered or segregated) phases are well estimated from calculated $\Delta{E}_{form}$'s:\cite{Zarkevich2007,Alam2010} For segregating CSAs ($\Delta{E}_{form}>$0), T$^{MG}_{c}\approx {\Delta{\text E}}_{form}/S_{pt}$; \cite{Zarkevich2007} and, for ordering CSAs with ${\Delta{\text E}}^{\alpha\rightarrow\beta}_{form}$= ${\Delta{\text E}}^{\alpha}_{form}-{\Delta{\text E}}^{\beta}_{form} >$ 0, the order-disorder transition  is T$^{od}_{c}$$ \approx$$ \Delta \text{E}^{\alpha\rightarrow\beta}_{form}$.\cite{Alam2010} Notably, \#4  also reveals if vibrations are important, as vibrational entropy in binaries correlates as $\Delta S_{vib}$ =$-\Delta\chi/3~(\pm0.06\Delta\chi)$.\cite{Fultz2004}  Thus, we may quickly estimate T$_c$ for $d$-band CSAs  as  $\text{T}^{\alpha\rightarrow\beta}_{c} \approx \text{T}^{\alpha\rightarrow\beta}_{c,pt}\left[ 1 + \Delta S^{\alpha\rightarrow\beta}_{vib}/\Delta S^{\alpha\rightarrow\beta}_{c,pt} \right]^{-1}$, which reproduces measured trends without phonon calculations.\cite{Zarkevich2007} These estimates are within $5-10\%$.\cite{Zarkevich2007,Alam2010}

As a predictive guide, we use KKR-CPA results to tune ($\Delta{E}_{form}$, $\delta$, \sVEC, and $\Delta\chi$)  versus $\{c_{\alpha}\}$ to find  ($\text{Mo}$-$\text{W}$)-$\text{Ta}$-($\text{Ti}$-$\text{Zr}$) alloys  in 5D space with better stability and mechanical properties. Results identify the stability of competing phases, possible multiphase regions, electronic properties, and practical design limits.  We use the above criteria to restrict search space for mechanical simulations.

\begin{figure}
\caption{(Top) Calculated formation energies ($\Delta{E}_{form}$) for Ta$_{x}$W$_{1-x}$ agree well with measured values. (Bottom) For Mo$_{x}$(WTaTiZr)$_{1-x}$, ($\Delta{E}_{form}$) versus x showing relative stability of A1, A2, and A3 phases. While A2 is most favored, positive $\Delta{E}_{form}$ indicates segregation tendency, so $\Delta{E}_{form}\approx$0 stabilizes the A2 phase.}
\includegraphics[scale=0.35]{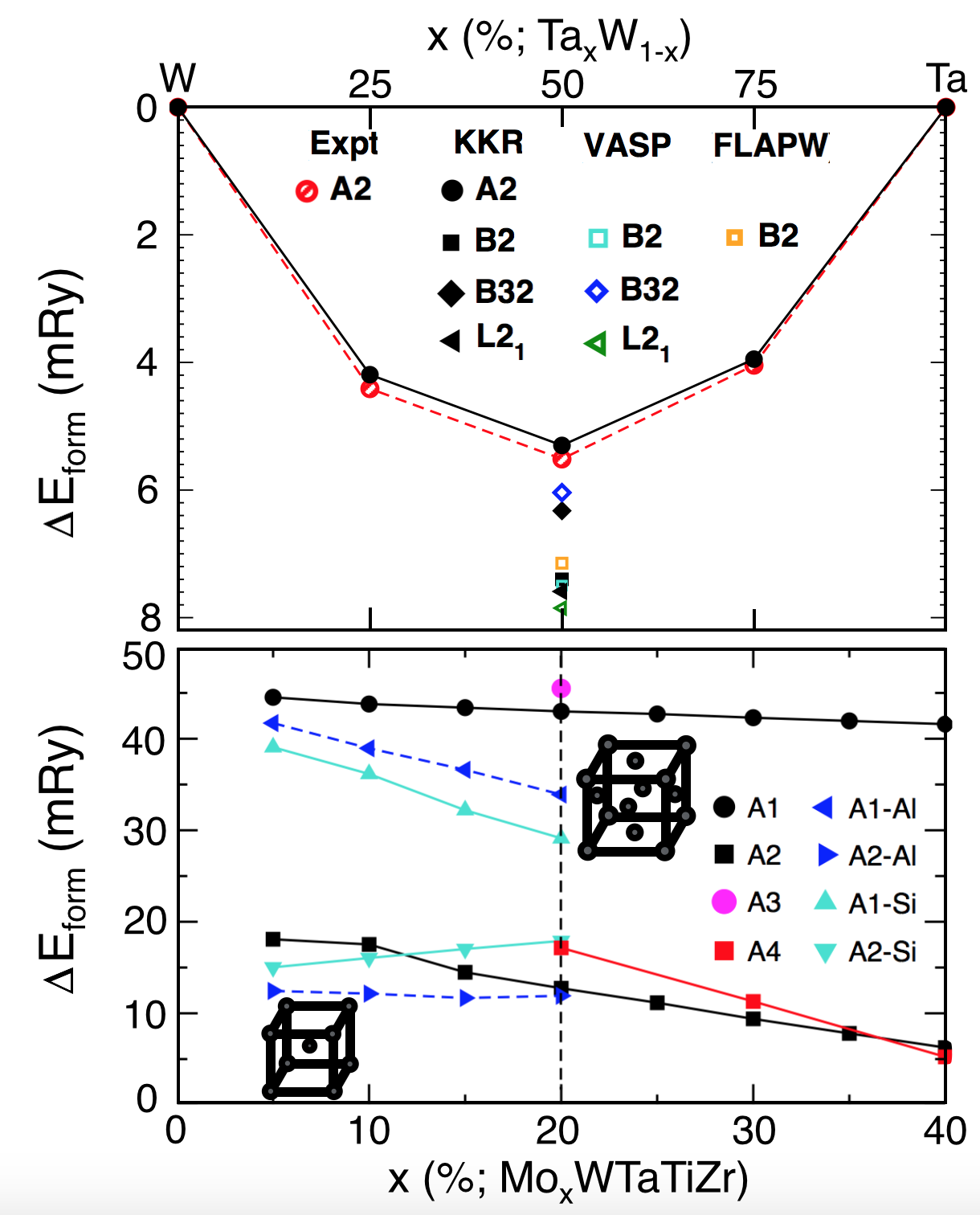}
\vspace{-0.3cm}
\label{1}
\end{figure}

\noindent   Here, via the KKR-CPA, we search all CSAs without restrictions on $\{c_{\alpha}\}$, or the need for large supercells, as A1,A2 (A3) have only 1 (2) atoms per cell. In this quinary, atomic size of Zr (1.60~\AA) is largest, followed by Ti, Ta, and W, Mo  (1.46, 1.43, and 1.37, 1.36~\AA), where bandwidths (inversely related to atomic size) and alloy hybridization determine the effect of size.\cite{Pinski1991} 
For $\chi$ (or $\Delta\chi$'s), reflecting solubility and vibrational entropy,\cite{Fultz2004,Zarkevich2007} (W,Mo) have largest $\chi$ (2.36, 2.16), followed by (Ti, Ta, Zr) with (1.54, 1.50, 1.33). From $\Delta\chi$ (Mo, W) would have the largest solubility (mixing) range, while \%Zr is smaller based on $\delta$. Larger \%W increases `E' for engineering needs, but increases weight, and \%Zr reduces Ti content while positively impacting flow stress.\cite{Flowstress}

\begin{figure*}[t]
{\caption{$\Delta{E}_{form}$~and~B (inset) vs.~($x, y$) for (MoW)$_x$Ta$_{y}$(TiZr)$_{1-x-y}$, a cut  through 5D $\{c_{\alpha}\}$ space containing equiatomic composition ({C1}), and plotted with vertical (dashed) lines of constant~\sVEC~(4-6) and contours of constant $\delta$ (1-6).  Besides global stability ($\Delta{E}_{form}$), alloys exhibit  local instability (SRO) to segregation (triangles) or ordering (circles). DFT results are every 5\% and interpolated.  Maximal B is near C7-10 with $\sim$300 GPa, where $\Delta{E}_{form}$$\sim$0 and segregation is lost (Fig.~{5}).}}\label{2}
{\includegraphics[scale=0.35]{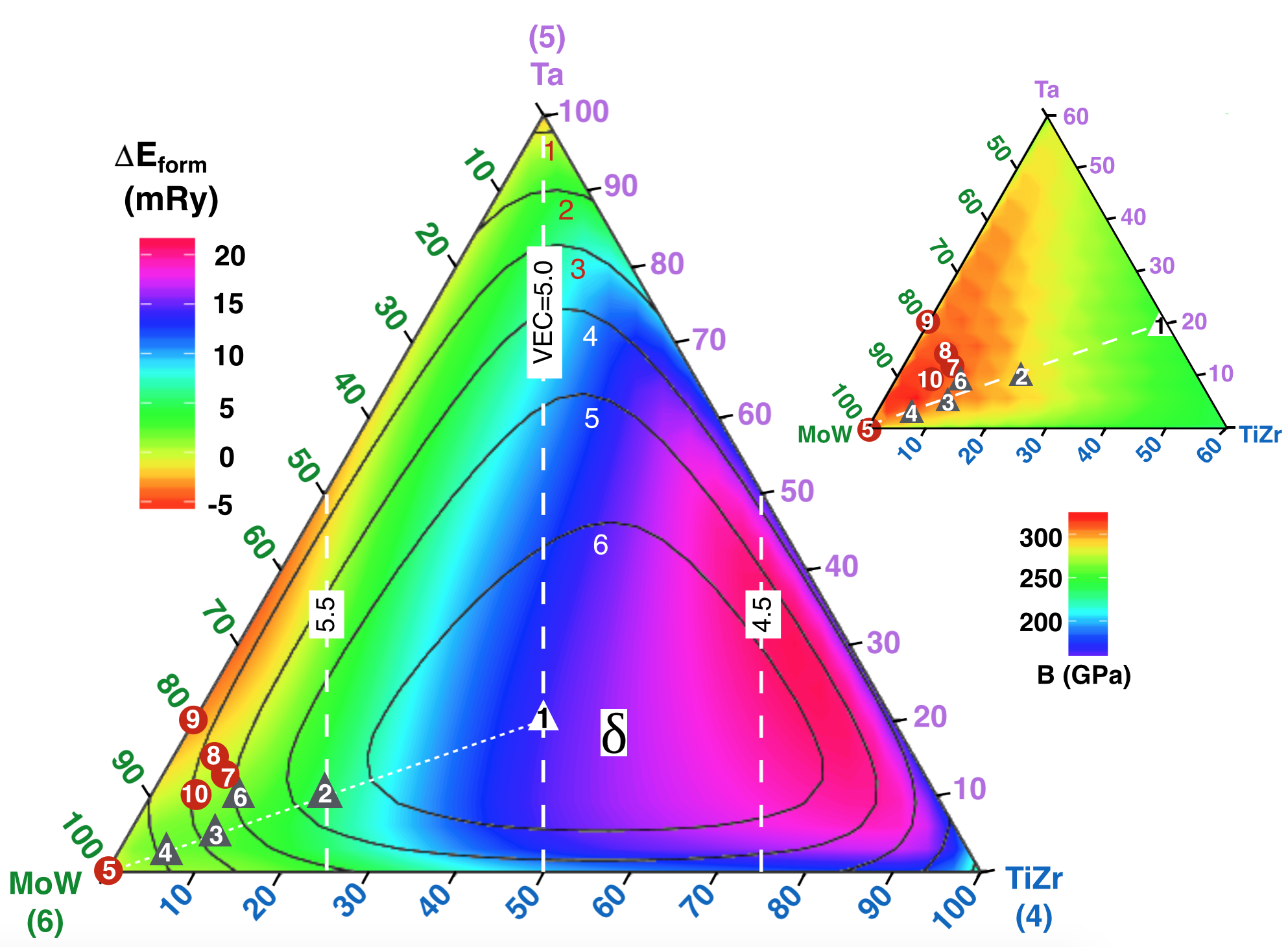}}
\end{figure*}

\noindent
\subsection{\bf Design \& Assessment:}~First, we  exemplify in Fig.~{1}, top-panel, our accuracy for $\Delta{E}_{form}$ vs. $x$ in Ta$_{1-x}$W$_x$, which agrees well with measured values (within 5\%), and ordering enthalpies are low ($<$310~K) compared to melting. Also, we show results for specific ordered cells, which are compared to and agree well with other reliable band-structure methods [e.g., VASP pseudo-potential \cite{VASP} and full-potential linear-augmented plane-wave (FLAPW) \cite{FLAPW}]. For Mo$_{x}$(WTaTiZr)$_{1-x}$ in Fig.~{1} (bottom-panel), we find that A2 is favored over A1 or A3, and that increasing \%Mo (larger $x$) helps stabilize A2. So, in this work, we focus on A2-phase of (Mo-W)-Ta-(Ti-Zr) alloy. For $x\gtrsim0.4$ A4 phase competes with A2, and Frank-Kasper phases, like C15-Mo$_{2}$(Ti-Zr), may be anticipated. Usually elements from group IIA/IVA of the periodic table, e.g., Al/Si, are added to stabilize or change, e.g., oxidation resistance. However, we find that adding Al stabilizes A2-phase up to 20\%Mo, and similar behavior of Al-addition has been seen in other alloys too.\cite{AS2016} On the other hand, Si-addition comes out to be energetically less favorable than Al-addition.

\noindent  {\small {\it High-Throughput Assessments}}: For ($\text{Mo}$-$\text{W}$)-$\text{Ta}$-($\text{Ti}$-$\text{Zr}$) results are most easily presented in a cut through 5D $\{c_{\alpha}\}$ space to visualize with only two parameters  (${x}$, ${y}$) along lines or planes (Fig.~{2}), changing  $\{c_{\alpha}\}$ in obvious ways. For fast screening of (Mo-W)-Ta-(Ti-Zr) 5D composition design space, we used an estimate for lattice constant to perform ``high-throughput" \cite{8,9,10,11,12,13,14} calculations to discover the best alloys in terms of phase-stability and/or mechanical behavior. Specifically, we estimated alloy lattice constants via Vegard's rule, which is the concentration-weighted sum of volume optimized elemental lattice constants in the parent alloys (A2) phase, or, simply, a$_{alloy}$ = $\sum_{i}\left[c_{i}^{X}a_{i}^{X}\right]$, where i=1, 5 and X= Mo, W, Ta, Ti, Zr. The estimated lattice constants are within 1-3\% with respect to the optimized lattice constants for all considered compositions. To down-select regions of interest, we perform the calculation over the entire design space and chose increments in $\{c_{\alpha}\}$ every 5\%  to sweep whole 5D space, see Fig.~{2}. For selected alloys, we perform full lattice-optimization to determine $\Delta{E}_{form}$ and B, and detail the electronic-structure (dispersion and density of states), and the thermodynamic short-range order (incipient ordering) for alloy design. 

\noindent   {\small {\it Assessment \& Validation}}: 
We now assess CSAs that best satisfy design criteria, and local stability. 
Along with other targets, KKR-CPA $\Delta{E}_{form}$ vs. $\{c_{\alpha}\}$ for (Mo-W)$_x$Ta$_{y}$(Ti-Zr)$_{1-x-y}$ are shown in Fig.~{2}.
Clearly, $\Delta{E}_{form}$ for equiatomic case is too positive ($+12.7$~mRy), and decomposition is expected (with T$^{MG}_{c}$=1244\,K from estimates in Hume-Rothery section). Our calculated SRO also indicates phase decomposition at spinodal T$_{sp}$=1240~K,  agreeing with T$^{MG}_{c}$, in the near-equiatomic and Ti-Zr-rich alloys (see Fig.~{5}). 
This predicted segregation is corroborated by our X-ray diffraction experiments, Fig.~{3}(a), that indicate presence of two (major/minor) phases. The phases were indexed as a disordered A2 phase with $Im{\bar3}m$ space group, and a minor phase of $Fd{\bar3}m$ space group. The A2 lattice parameter was measured as 3.1713 $\AA$ (std. dev: 0.0002 $\AA$). Figure~{3}(b) shows the SEM micrograph of the alloy with a two-phase alloy evident. The major phase (A2) is Mo, W and Ta rich, with small amounts of Ti and Zr incorporated in it. Given the higher melting temperatures of Mo, W and Ta, the major phase is likely to be the primary solidifying phase during the final step of casting. As this phase forms during casting, Ti and Zr are rejected into the surrounding liquid, which subsequently freezes. Hence, the minor phase is Ti and Zr rich and incorporates small amounts of the refractory metals. 

\begin{figure}
\caption {(a) X-Ray diffraction pattern from the equiatomic alloy; (b) Backscattered SEM of the equiatomic alloy $-$ the brighter contrast is the (Mo,W,Ta)-based solid solution, while the darker phase is the (Ti,Zr)-rich phase. The scale bar is 300 $\mu{m}$. Composition in major-phase ($Im-3m$) of  Mo/W/Ta/Ti/Zr is 22.96 (1.79)/33.63 (2.57)/28.31(0.44)/11.19 (1.79)/3.90 (0.52); and composition in minor-phase ($Fd(-3)m$) of Mo/W/Ta/Ti/Zr is 8.44 (0.82)/0.89 (0.15)/3.06 (0.22)/32.61 (0.35)/55.00 (0.87). Phase compositions, as determined using EDS. Compositions are provided in at.\%. Standard deviations are provided in parentheses.}
\includegraphics[scale=0.35]{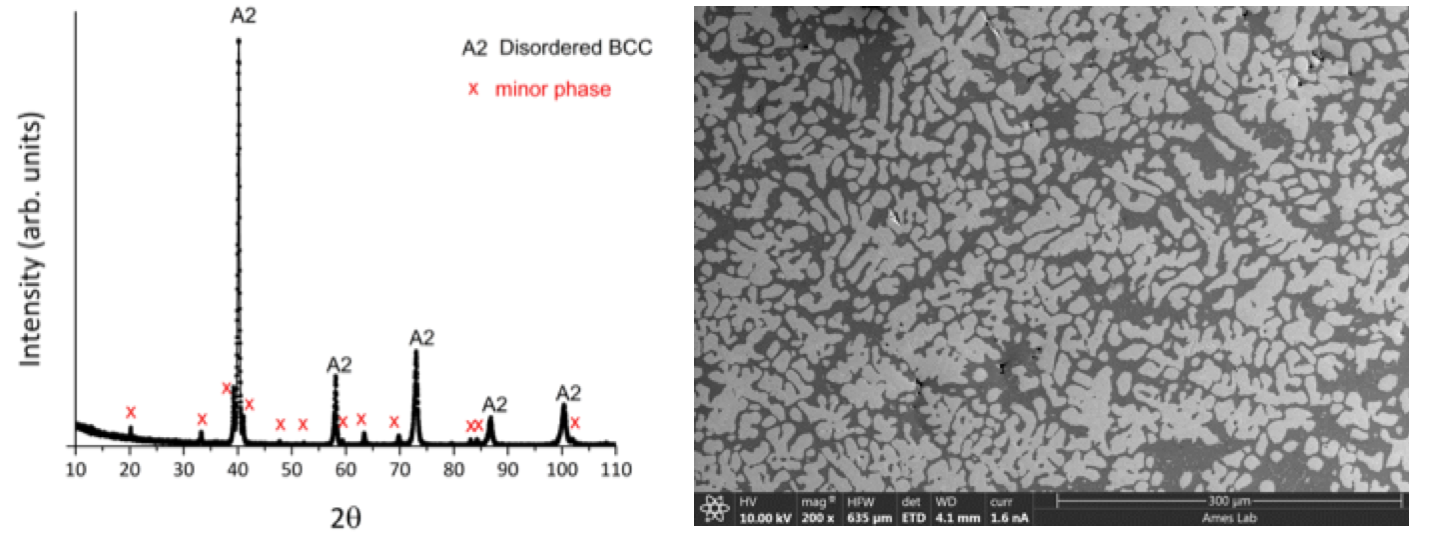}
\label{3}
\end{figure}

\noindent   To visualize key alloying effects for these CSAs, we plot in Fig.~{4} the electronic dispersion and projected total density of states (TDOS), referenced to each alloy's Fermi energy, E$_{F}$. With disorder, dispersion exhibits broadening in E and ${\bf k}$, showing that ${\bf k}$ is a ``good'' (on the scale of the Brillouin zone) but not an exact quantum number (as for zero-width, ordered bands); the width $d{\bf k}\sim l_e^{-1}$  ($l_e$ is the electron scattering length) and gives rise to increased residual resistivity, as may be calculated.\cite{WHB1985}

\noindent   Guided by such details, we can improve CSA properties. The equiatomic alloy has a TDOS with E$_{F}$ not yet in the pseudogap between bonding and antibonding states (top Fig.~{4}), so this alloy does not satisfy the design criteria. The VEC (the average electrons per atom outside the closed shells of the component atoms) is a dominant factor in controlling the phase stability of the alloys. The electronic states present on/near the E$_{F}$ are chemically most active, which affect the chemical property of the alloy, i.e., more states at  E$_{F}$ destabilizes the alloy. This means, for such cases, adding or removing electron we can manipulate the electronic properties very quickly. By integrating states from E$_{F}$ to the pseudogap for equiatomi case, $0.2$ electrons are needed to fill bonding states and improved stability. More at.\%Mo-W (\sVEC=6) adds electrons, moving E$_{F}$ up (Fig.~{4}), and $\Delta{E}_{form}$ reduces to stabilize CSAs (Fig.~{2}).
Adding small \%Ta helps in altering states near E$_{F}$: the flat bands near $\Gamma$ in Fig.~{4} are moves from E$_{F}$ for C7, lowering $\Delta{E}_{form}$, shown in supplementary figure 2 \& 3. We show in Fig.~{5} that the SRO changes from clustering in C6 to ordering in C7, while $\Delta{E}_{form}$$\sim$0.
Here, $\Delta{E}_{form}$ reduces quickly for MoW with a small \%Ta and \%TiZr, while bulk modulus (B) increases quickly (inset Fig.~{2}).
Notably, the dispersion of A2-metals is canonical when scaled by bandwidth (inverse atomic size), and so the
behavior of the alloy dispersion is fairly generic and predominantly determined by relative composition and size, hybridization, and band filling.

\noindent   To promote oxide-scale formation for protection, and light-weighting, Al is often added. In Fig.~{4}, 5\%Al added at the expense of Ta to C6 (whose $\Delta{E}_{form}$=+0.70~mRy) increases disorder broadening (from Al $sp$-$d$ hybridization) and causes $d$-state around $\Gamma$ (predominantly TiZr) to again straddle at E$_{F}$. This Fermi surface feature energetically destabilizes the alloy making $\Delta{E}_{form}$ much more positive (+6.8~mRy, with T$^{MG}_c$=828~K), which is also visible in SRO showing strong clustering behavior at T$_{sp}$=780~K (see Supplementary Figure 4). The most significant Al-Mo pair suggest that Al will segregate to surfaces due its faster kinetics, as needed for oxide formation, i.e., adding Al at the expense of Ta or TiZr decreases VEC and drops E$_{F}$ into localized $d$-states, reduces stability; so a balance must be struck by keeping some Ta and TiZr and making VEC high enough to be near $\Delta{E}_{form}$$\sim$0 but with a large B (Fig.~{2}). Unlike in other systems, Al  is not generically a good A2 stabilizer, as it leads to larger electron scattering for reduced stability, increased resistivity, and decreased thermal transport, see [\onlinecite{PS2015}] and references therein.

\begin{figure}
\caption {For {C1}, C6, C7 in Fig.~{2}, the electronic dispersion along Brillouin Zone high-symmetry lines, and the projected total density of states (TDOS). Effects of Al doping is also shown (bottom). By tuning $\{c_{\alpha}\}$ to add electrons, shift dispersion, or change disorder, the pseudogap is reached near E$_{F}$ ($0$), better stabilizing the alloy and altering segregation.}
\includegraphics[scale=0.35]{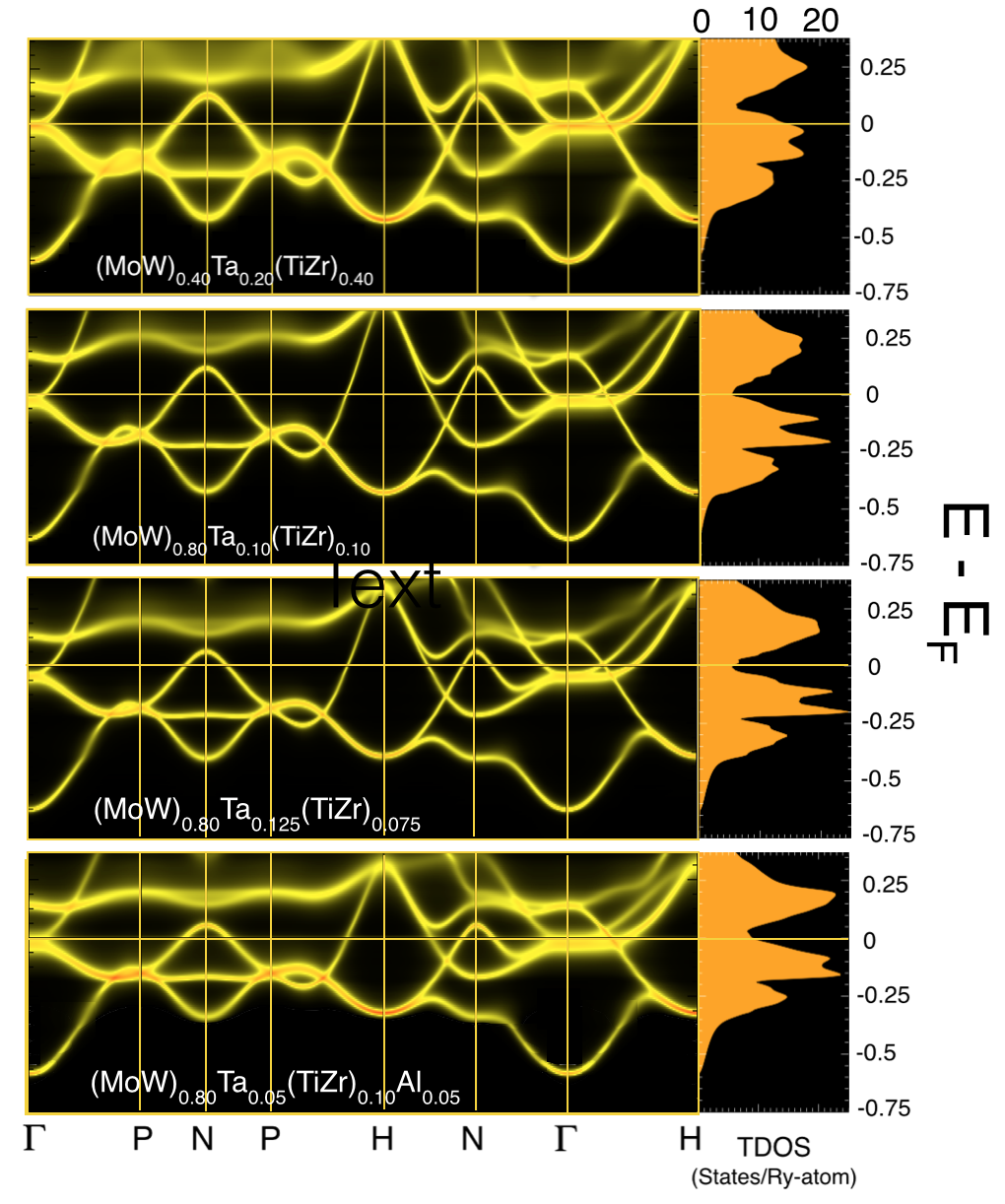}
\label{4}
\end{figure}

\begin{figure*}
\caption {For three alloys in Fig.~{2}, SRO correlations (top)  $\alpha_{\mu\nu}$({\bf{k}};1.15T$_{sp}$) and (bottom) energies S$^{(2)}_{\mu\nu}$({\bf{k}};1.15T$_{sp}$) plotted along high-symmetry lines in the Brillouin zone: (left) C1  ($x$=$2/5$, $y$=$x/2$) with $\alpha_{TiZr}$($\Gamma$) clustering with T$_{sp}$=1240~K; (middle) C6 ($x$=4/5,$y$=$1/10$)  with $\alpha_{ZrMo}$($\Gamma$) clustering with T$_{sp}$=500~K; and (right)  C7 ($x$=$4/5$, $y$=$1/8$) with $\alpha_{TiMo}({\bf k}_{0})$ with incommensurate ordering  $|{\bf k}_{0}|=0.7|$N-H$|$  with T$_{sp}$=298~K. For C1 [C6], Ti-Zr [Zr-Mo] pairs dominate correlations, but clustering is driven by S$^{(2)}_{\text{ZrMo}}$($\Gamma$). In C7, Ti-Mo dominate correlations, but Ti-Mo and Ta-W drive ordering.}
\includegraphics[scale=0.35]{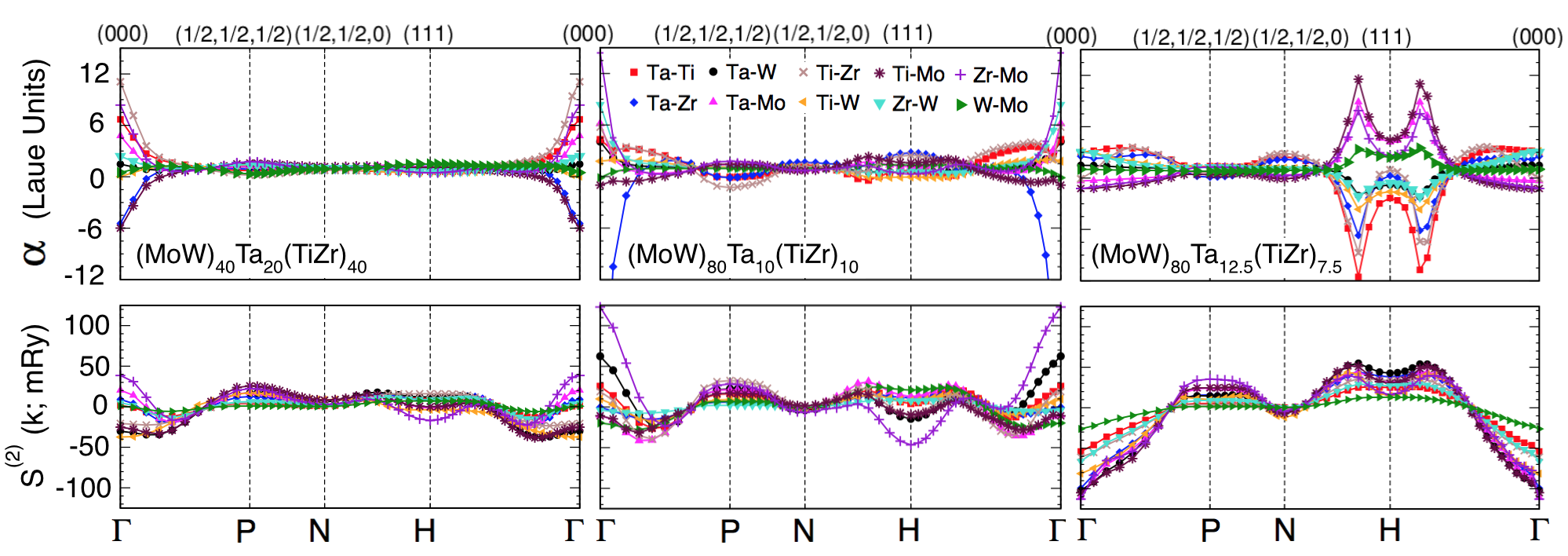}
\label{5}
\end{figure*}

\subsection{Chemical SRO:} From KKR-CPA linear response (see Methods), we predict  (Fig.~{5}) Warren-Cowley SRO (or atomic pair correlations)  $\alpha_{\mu\nu}$({\bf{k}};T), whose largest peak at wavevector {\bf k}$_{0}$ reveals the unstable (Fourier) modes to ordering, or clustering at $\Gamma$=(000).
As an alloying guide, SRO identifies pairs driving the instability, and predicts the spinodal T$_{sp}$, where $\alpha^{-1}_{\mu\nu}({\bf k}_{0};T_{sp})$=0 signifying the absolute instability to this chemical fluctuation.\cite{PS2015,AS2016} 
In real-space, pair probabilities are  $P^{\mu\nu}_{ij}= c_{\mu}^{i} c_{\nu}^{j} (1-\alpha_{\mu\nu}^{ij})$, with $\alpha_{\mu\nu}^{i \neq j}$=$0$ for no SRO, and $\alpha$$<$0 ($\alpha$$>$0) indicates ordering (clustering) with bounds of   $-{[\min(c_{\mu},c_{\nu})]^2}{(c_{\mu}c_{\nu})^{-1}} \leq \alpha_{\mu\nu}^{i\neq j} \leq 1$. 

\noindent   Near-equiatomic alloys in Fig.~{5} have maximal SRO peaks in $\alpha_{\mu\nu}$({\bf k}$_{0}$=$\Gamma$;T$>$T$_{sp}$) signaling spinodal (infinite wavelength) decomposition in specified pairs at T$_{sp}$ of 1240 K for C1, and at 500 K for C6. This 60\% drop in T$_{sp}$ is unsurprising given that $\Delta{E}_{form}$ reduces with Mo-W and Ta addition (Fig.~{2}). 
For C7, where $\Delta{E}_{form}$ has become slightly negative due to movement of bands present at $\Gamma$ away from E$_{F}$, a weak incommensurate (long-period) ordering is found with SRO peak (Fig.~{5}) at 70\% along N--H at {\bf k}$_{0}$=$(0.85,0.15,0)$.
This SRO arises from Fermi-surface nesting,\cite{AJP1995} with contributions at a radius of $|{\bf k}_{0}|$$\sim$$0.86$, as confirmed along $\Gamma$-$\text{H}$ (Fig.~{5}).  [SRO is B2 type if it peaks at ${\bf k}_0$=H=$\{100\}$, commensurate with A2 lattice.] For theory and detailed examples, see [\onlinecite{PS2015}] and [\onlinecite{AS2016}]. We also show, in supplementary figure~4, that 5\%Al addition to the C6 alloy instigates a clustering instability. The Al-Mo pair drives spinodal decomposition at T$_{sp}$ of 780 K, which shows the tendency of Al to phase separate from Mo, an indication that Al's clustering tendency might be helpful in promoting stable oxide-layer at high temperatures. These results indicate that alloying may improve oxidation behavior, just as for Fe-Cr with a narrow window for chromia formation. Clearly, KKR-CPA methods address profound electronic and alloying effects not possible from effective potentials, or methods that approximate disorder by ordered configurations.

\begin{figure*}
\caption {(a, bottom) Simulated stress-strain for uniaxial quasi-static compression at 300 K on ideal crystals for equiatomic C1 alloy, which exhibits simple plastic flow, and for C4 alloy, where stress drops arise from dislocation motion and annihilation. (a, top) Stress-strain versus T  from 77 to 1100 K for C4 alloy.  (b) C4 strained elastically to 0.065 (b-1, blue), then, at yield, shear-bands appear (b-2, black) accompanied by dislocations.  At high strain-rates ($\sim$0.3), stronger edge (red) and screw (blue) components are found (b-4) with Burgers vector (green). (c) For (non)equiatomic alloys, E~vs~T from classical MD [diamonds], KKR-CPA results [squares, extended to 600 K via Gr{\"u}niesen theory, then extrapolated (dotted lines)], and measurements (Expt) for W, Mo, commercial TZM,\cite{TZM} and C1 alloy (present experiments). (d) For validation in A2 Mo$_{z}$W$_{1-z}$, measured (black symbols) and KKR-CPA results (blue lines and circles) for B, E (GPa), lattice constant $a$ (\AA), and density $\rho$ (Mg/m$^3$), with Poisson ratio ($\nu$) from MD. Also, predicted values are shown for quinary (Mo$_{z}$W$_{1-z}$)$_{0.85}$Ta$_{0.10}$(TiZr)$_{0.05}$ [red lines and squares],  where C10 ($z$=0.5) response is similar to C4 in Fig.~{6}c.  {\bf Design region:} Region around quinary $z$=0.05 (denoted C$\bar{10}$, at Mo-rich end of a line perpendicular to plane at C10 in Fig.~{2}) shows enhanced E (above Mo) and better T-dependent slope (Fig.~{6}c) from a favorable Poisson effect (green area in Fig.~{6}d). Above 500 K, C$\bar{10}$ has larger, less T-dependent E than TZM (2.3$\times$ at 2000 K).}
\includegraphics[scale=0.35]{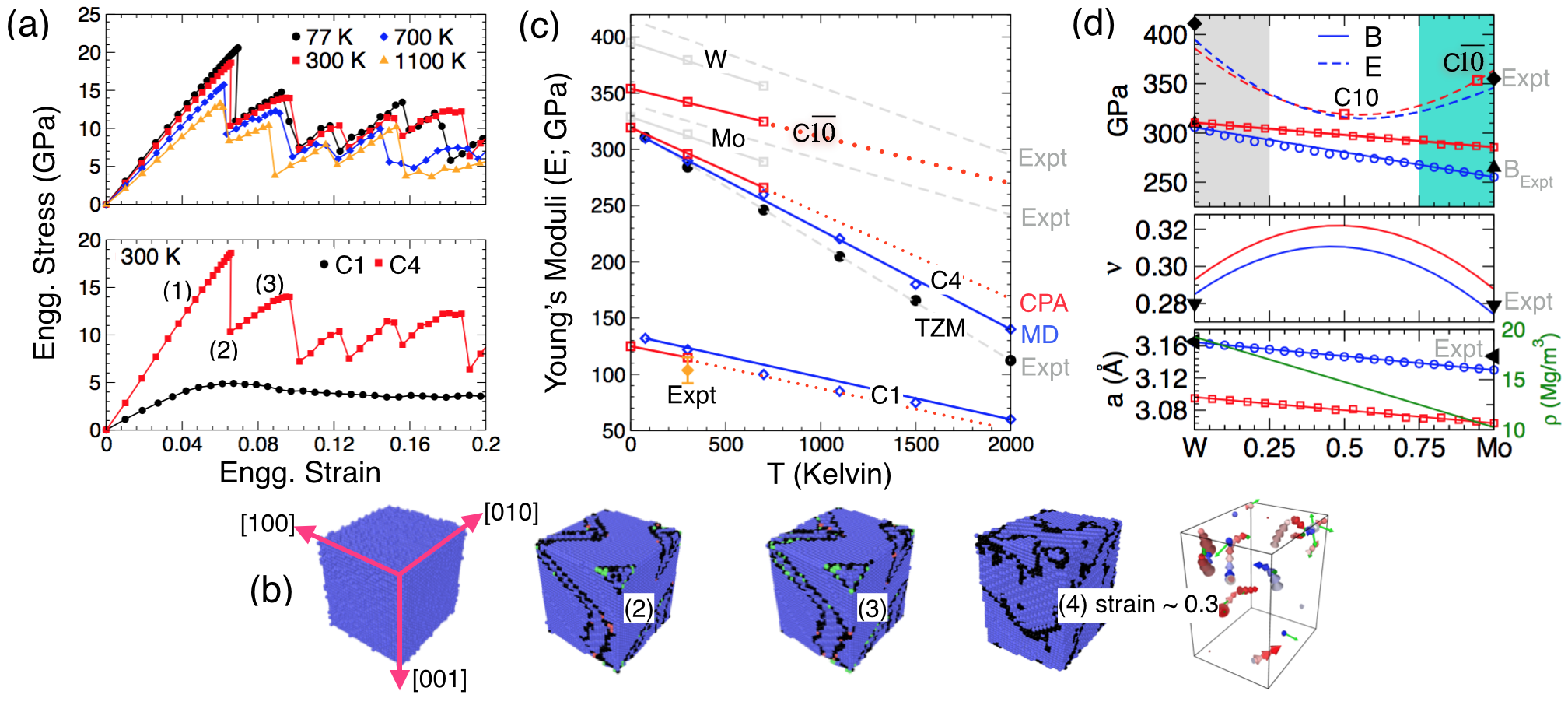}
\label{6}
\end{figure*}

\noindent   \subsection{Deformation Analysis:} Mechanical properties in CSAs have been studied at macro- and micro-scopic levels,\cite{AMG2016, YZ2014} but deformation analysis is key to establish high-T structural candidates. We perform quasi-static uniaxial loading via MD simulations (see Methods) by deforming an ideal single-crystal alloy in small but finite steps and equilibrating after each step. 
For equiatomic case, $<$100$>$ compression (Fig.~{6}a) reveals a smooth stress-strain curve signaling simple plastic flow. In contrast, C3 (0.425at.\%Mo) ideal crystal has stress drops and strain-hardening triggered by $<$111$>$ dislocations;  a stress drop at $0.065$ strain marks the initiation of dislocation with  A2 Burgers vector, {\bf b} = $\frac{1}{2}$$<$111$>$, from 77--2000 K. 

\noindent     Snapshots of the evolution show that dislocations (edge and screw type) triggered these instabilities (Fig.~{6}b). 
The defect mobility is affected by local distortions caused by the different sizes and modulus of the solutes. The rise and drop in stress with increasing strain in the ideal crystal corresponds to the defect evolution where new dislocations occur after every major stress drop followed by strain-hardening due to dislocation interactions and drag.
An investigation of the local structural environment (Fig.~{6}b) reveals deviation from perfect A2, as yielding occurs for 300 K.  Shear bands (black) are promoted, denoting deformed regions with higher compression. At very high strain the interplay between edge and screw dislocations can be visualized via the band dynamics (Fig.~{6}b-4). 

\noindent     For engineering, Young's modulus E =3(1-2$\nu$)B is pertinent, so Poisson's ratio $\nu$ is also key. Small-deformation MD simulations determined E and $\nu$ at 300 K, and E~vs.~T was found from the elastic stress-strain curve (Fig.~{6}a). The KKR-CPA energy versus $a$ (A2 lattice constants) at $\{c_{\alpha}\}$ determines the equilibrium $a_0$, $\Delta{E}_{form}$, and B (Fig.~{6}d), all used in Fig.~{2}. 
We compare temperature dependence of E, in Fig.~{6}c, calculated from  MD, KKR-CPA (using Gr\"uneisen approximation at low T), and experiments for commercial Mo-rich TZM alloy,\cite{TZM} which again validate theory results. As MD is performed on ideal crystals, an ``ideal'' yield strength is obtained, with a qualitative relative change versus temperature.  

\noindent    To confirm our predicted E in equiatomic C1, we performed indentation on samples prepared by arc-melting (see Methods), in Fig.~{6}(c), and show that value from measurement 104$\pm$12 GPa at 300 K compared very well to our predictions $115$ GPa from KKR-CPA and $120$ GPa from MD.
As B changes slowly for binary Mo-W (Fig.~{6}d), the Poisson effect (variation of $\nu$) controls strength, which requires W- or Mo-rich alloys for larger E values (Fig.~{6}d). Similarly, for quinary (Mo$_{z}$W$_{1-z}$)$_{0.85}$Ta$_{0.10}$(TiZr)$_{0.05}$, we find that C10 ($z=0.50$) has strength similar to C4 (Fig.~{6}c). Whereas we predict a region around C$\bar{10}$ ($z=0.05$, highlighted in Fig.~{6}d, which is perpendicular to plane in Fig.~{2} at C10) that shows both enhanced stability and E (Fig.~{6}c). For these CSAs, we find $3\times$ larger E than high-entropy alloys at 300 K, and alloys like C$\bar{10}$ have a much larger, less temperature-dependent modulus (Fig.~{6}c) above 500~K ($2.3\times$ at 2000 K) over existing commercial TZM alloys, and lie midway between pure Mo and W, unlike TZM alloys.

\noindent   Finally, one notable point, albeit not surprising, the classical MD simulations fail to represent properly the alloys that crossover from $\Delta{E}_{form}$ positive to negative (e.g., C4-to-C7 or C3-to-C10) in which electronic dispersion (not addressed by semi-empirical potentials) is controlling the materials physics. Hence, we plot only C$\bar{10}$  KKR-CPA values in Fig.~{6}c, as MD values of E~vs.~T  for C10 and C$\bar{10}$ are similar to C4, whereas C$\bar{10}$ values from first-principles increases over C10, as expected from Fig.~{6}(d).  

\noindent  From a design perspective, in general, alloys in complex solid-solution alloys have superior properties over near-equiatomic alloys (so-called high-entropy alloys), although the design space becomes enormous. 
Using a first-principles KKR-CPA, we predicted the relative phase stability, dispersion, short-range order (i.e., incipient long-range order, including T$_c$) and its electronic origin, and mechanical properties over all compositions as a design guide.
Using electronic alloy design concepts and criteria, we identified higher strength refractory (Mo-W)-Ta-(Ti-Zr) alloys from materials physics and engineering perspectives.
Temperature-dependent deformation (most relevant the elastic behavior) in selected set of alloys was modeled using classical MD simulations, validated from first-principles data; we also identified failures in classical potentials that arose from dispersion effects.

\noindent     Based on our calculation, we designed a Mo-rich region of improved stability with enhanced Young's moduli over high-entropy alloys, as we confirmed experimentally, and an improved temperature-dependence above 500~K  ($2.3\times$ at 2000 K) over existing commercial alloys.  Our electronic-structure  approaches and analysis of alloying and stability (formation energies, dispersion, short-range ordering) highlights how instructive these details are in guiding design. The techniques are quite general for assessing any arbitrary complex solid-solution alloys, where alloying and non-trivial electronic effects play a key role.

{\small 
\section*{Methods} }
\vspace{-1em}

{\small 
\subsection {DFT Methods:} KKR electronic-structure is used with the coherent-potential approximation (CPA) to handle chemical disorder; \cite{JohnsonCPA, JP1993} screened-CPA addresses Friedel screening from charge-correlations.\cite{JP1993} Scalar-relativistic effects are included (no spin-orbit). Generalized gradient approximation to exchange-correlation was included through use of \texttt{libXC} libraries.\cite{libxc2012} CSAs require only 1-atom (2-atom) cells for A1,A2 (A3). Brillouin zone (BZ) integrations were performed with Monkhorst-Pack  k-point method,\cite{Monkhorst} with $12\times12\times12(6)$ for A1,A2 (A3) meshes. We used 300 $\it{k}$-points in the irreducible-BZ to visualize  dispersion along symmetry lines. Each scatterer's radii were defined by neutral ``atoms-in-cell'', with interstitial divided  proportionally to each scatterer, to improve radial density representation near saddle-points in the electronic density.\cite{AJ2009,MECCAv2} We chose $L_{max}$=3 spherical-harmonic basis to include {\emph{s, p, d}} and {\emph{f-}}orbital symmetries. Shallow core states were included in the valence in all calculations. A variational potential zero $\text{v}_0$ was used to yield kinetic energies nearing those of full-potential methods.\cite{AJ2012} For self-consistent densities, complex-energy contour integration\cite{J1985} used 20-point Gauss-Legendre semicircular contour.
}

{\small \subsection {Chemical SRO:} From KKR-CPA linear-response, we calculate SRO parameters, $\alpha_{\mu\nu}$({\bf k};T), for $\mu$-$\nu$ pairs,\cite{PS2015,AS2016}  as detailed elsewhere.\cite{AJP1995,Pinski1991,AJPS1996,SJP1990,SJP1994}   
Dominant pairs driving SRO are identified from pair-interchange energies,  $S^{(i,j)}_{\mu\nu} (T)$, or curvature (concentration 2$^{nd}$-variation) of the KKR-CPA grand potential, yielding energy cost for concomitant fluctuations of $c^{i}_{\mu}$, $c^{j}_{\nu}$ at atomic sites $i$, $j$.   
$S^{(2)}_{\mu\nu} ({\bf k};T)$ reveals the unstable (Fourier) modes with ordering wavevector {\bf k}$_{0}$ (or clustering at (000)), identifies the origin for phase transitions, and dictates the SRO:  
$\alpha^{-1}_{\mu\nu}({\bf k};T)$\,=\,$[c_{\mu}(\delta_{\mu\nu}-c_{\nu})]^{-1}  [(\delta_{\mu\nu}c^{-1}_\mu  + c^{-1}_N) -  {(k_B T)^{-1}} S^{(2)}_{\mu\nu} ({\bf k};T)]$. The spinodal temperature, where $\alpha^{-1}_{\mu\nu}({\bf k}_{0};T_{sp})$=0, signifies an absolute instability to this fluctuation and provides an estimate for T$^{MG}_c$ or T$^{od}_c$.\cite{PS2015,AS2016,PS2017}   For N$>$2, pairs driving ordering (clustering) will not necessarily be the same pairs that peak in the SRO due to the matrix inversion that relates them (Fig.~{5}).}

{\small \subsection {MD Simulations:} Deformation is evaluated using Large-scale Atomistic/Molecular Massively Parallel Simulator (LAMMPS) package. \cite{Plimton1995} The KKR-CPA structural parameters are used to validate potentials for finite-T modeling. The force-field parameters for the quinary are established from available ternary EAM potentials.\cite{Zhou2004} 
We verified similar hybrid potential parameter combinatorial technique for high-entropy aloy, like Al$_{10}$CrCoFeNi.\cite{AS2016} 
In A2 lattice, with dimensions (30$\times$30$\times$30){\it a}  (54,000 atoms), we distributed 5 elements via composition to form (Mo-W)-Ta-(Ti-Zr) solid solutions. Initially the lattice was melted at 4,000 K for 90 ps, followed by a quench to 300 K within 10 ns. Uniaxial deformation was performed after equilibrating and relaxing the structure at high strain, as detailed in supplementary video 1.\cite{AS2016}}

{\small \subsection {Synthesis \& Characterization:} The equiatomic MoWTaTiZr was synthesized by arc-melting pellets of elemental powder blends (Alfa Aesar, purity $\geq$ 99.9\%) in an ultra-high purity argon atmosphere on water-cooled copper hearth. Powders were used to reduce the large macro-segregation that occurs during casting when using elemental chips. With the  significant difference in melting temperatures (3695 K for W vs. 1941 K for Ti)  a three-step melting process was adopted. Step 1: W and Ta powders were mixed thoroughly in a SPEX 8000 mill, and pressed using a Carver hydraulic press; and the pellet was then arc-melted. Step 2: elemental blends of Mo, Ti and Zr were similarly mixed, pressed and arc-melted. Step 3: both arc-melted buttons were re-melted together for a total of four times to ensure better homogeneity. }

\noindent    {\small Phase analyses were carried out using a Philips PANalytical X-Ray Diffractometer (XRD), in a Bragg-Brentano geometry using Cu-K$\alpha$ radiation. Microstructure and phase compositions were analyzed using a FEI Helios NanoLab G3UC Scanning Electron Microscope (SEM), equipped with Oxford Energy Dispersive Spectroscopy (EDS) system. Accelerating voltages of 10-15 kV were employed for imaging and compositional analyses. Compositions were measured at 7 different locations for each phase, with the average composition and X-ray diffraction shown in Fig.~\ref{3}. The diffraction pattern indicated the presence of two phases, indexed as a bcc (A2) phase with lattice parameter $3.1713(2)$~\AA~and a minor phase with $Fd{\bar 3}m$ space group, like for B32 (NaTl prototype) or C15 (MgCu$_2$ prototype) structures, with lattice constant $7.6148(9)$~\AA.}

{\small \subsection {Nanoindentation:} Nanoindentation utilized a tribo-indenter HYSITRON TI-900 with a Berkovich ($3\,\mu$m) tip. 
The indenter control module applies a trapezoidal load on the sample for $10\,s$,  followed by $5\,s$ rest, and unloads in $10\,s$. 
 To calibrate the sample measurements, which also determines the best applied load and optimum contact depth, the alloy was scanned (15 measurements) on an arbitrarily chosen sample location to optimize for force vs. displacement. For a minimum of $200\,n$m of contact depth, a $6000\,\mu$N load was found to  suffice. With these set, we indented the alloy at 30 manually chosen locations to measure the sample's elastic response; for the equiatomic case,  the mean values were:  Young's modulus of $103.73\pm11.49$ GPa,  hardness of $4.6\pm0.34$ GPa, and contact depth of $231.18\pm 8.74$ nm.}

\section*{ACKNOWLEDGEMENTS} 
Work supported by the U.S. Department of Energy (DOE), Office of Science, Basic Energy Sciences, Materials Science \& Engineering Division for theory/code development, and by the Office of Fossil Energy, Cross-cutting Research for application and validation for specific HEAs. Research was performed at Iowa State University and Ames Laboratory, which is operated by ISU for the U.S. DOE under contract DE-AC02-07CH11358. Work by AS, MSD, \& GB supported by the Office of Naval Research (grant N00014-16-1-2548), with computing resources from the Department of Defense High-Performance Computing Modernization Program.
\section*{Correspondence}
Correspondence should be addressed to PS~(prashant@ameslab.gov) and DDJ (ddj@ameslab.gov).

\end{document}